# Enhancing Physics Learning with ChatGPT, Bing Chat, and Bard as Agents-to-Think-With: A Comparative Case Study

Renato P. dos Santos
CIAGE – Centre for Generative Artificial Intelligence in Cognition and Education
ULBRA – Lutheran University of Brazil
E-mail: RenatoPSantos@ulbra.edu.br

**Abstract**

The rise of artificial intelligence (AI) has led to remarkable advancements in various fields, including education. AI models have shown their ability to analyse and provide instructive solutions to complex problems, making them attractive as pedagogical tools. This study aimed to compare and analyse the responses of four Generative AI-powered chatbots (GenAIbots) - ChatGPT-3.5, ChatGPT-4, Bing Chat, and Bard - within the constructivist theoretical framework. The assessment focused on Physics problems, viewing the GenAIbots as 'agents-to-think-with' and evaluating their tutor characteristics. Extensive interaction logs between the GenAIbots and a simulated student persona in simulated Physics learning scenarios were analysed using a single-case study methodology. The GenAIbots were presented with conceptually dense Physics problems to elicit insightful responses and promote deep understanding. Responses were qualitatively analysed based on seven tutor traits: subject-matter knowledge, empathy, assessment emphasis, cooperative environment creation, facilitation skills, comprehension of the learning process, and content knowledge-student facilitation balance. The findings revealed that all GenAIbots could function as agents-to-think-with, fostering critical thinking, problem-solving, concept comprehension, personalised learning, and displaying subject-matter knowledge. However, significant disparities were observed. ChatGPT-4 consistently demonstrated empathy and a deep understanding of the learning process, providing comprehensive, accurate, and contextually nuanced responses. While each GenAIbot had strengths, they also exhibited inconsistencies and shortcomings in their problem-solving approaches, emphasising the need for human intervention in AI-assisted learning. In conclusion, the study posits that while GenAIbots have limitations, their ability to function as agents-to-think-with in Physics education opens up promising prospects to revolutionise Physics instruction.

**Introduction**

According to Rivaldo, Taqwa, Zainuddin, and Faizah (2020), students face various difficulties in learning Physics. Identifying students' difficulties is crucial for designing effective Physics education. For Oladejo et al. (2023), those difficulties include mathematical concepts, abstract thinking, and problem-solving skills. Additionally, the authors suggest that technology can effectively enhance by providing interactive simulations and visual aids.

Language models have increasingly gained popularity in recent years due to their capacity to simulate human conversation and generate relevant responses (Borji, 2023). Among them is ChatGPT, a model developed by OpenAI. As outlined by Adiguzel et al. (2023), this model has been trained on vast quantities of data and has shown considerable proficiency in summarising academic articles, providing accurate solutions to issues, and performing more intricate natural language processing tasks with fluent and detailed responses.

The success and sophistication of ChatGPT have astonished global observers, including its creators at OpenAI, as it amassed over a million subscribers within a week of its public release on November 30, 2022 (Baidoo-Anu & Owusu Ansah, 2023).

Leike (2023) explains that the technology behind ChatGPT is not new; rather, it is an updated version of GPT-3, released in 2020. Nonetheless, the remarkable success of ChatGPT has spurred OpenAI to update and refine the chatbot since its debut persistently. The latest iteration, ChatGPT-4, benefits from training on an expanded and enhancing its natural language processing capabilities. Zhang's study (2023) provides a forward-looking perspective on ChatGPT's potential evolution and argues that while the launch of GPT-4 represents a minor leap in the field of Generative AI (GAI), it marks a significant step towards Artificial General Intelligence (AGI).

Given the staggering number of over 13,000 entries in Google Scholar that either feature ChatGPT in their titles or reference it in their abstracts, it is evident that there is an urgent need for academic studies focused on investigating the effective and productive use of Generative Artificial Intelligence-powered chatbots (GenAIbots), such as OpenAI's ChatGPT, Microsoft's Bing Chat, and Google' Bard in learning and research contexts.

The GenAIbots' proficiency in executing intricate tasks captivated global observers and stirred diverse reactions within the educational sphere. Many educators were intrigued by its potential to reshape traditional teaching practices, thereby making the learning experience more transformative (Adiguzel et al., 2023; Stokel-Walker, 2022). This proliferation of GenAIbots, with their advanced natural language processing capabilities, necessitates a reconsideration of established educational methodologies and assessment practices (Leike, 2023; Zhang, 2023).

The transformative potential of this generative AI tool to traditional teaching and learning methods has elicited diverse responses among educators (Adiguzel et al., 2023). A surge in its adoption within college teaching has been observed (Chun, 2023), where its implementation ranges across institutions. Some universities have issued guidelines to their professors on utilising a more diverse dataset ChatGPT, while others have granted instructors the liberty to make individual decisions. The potential of GenAIbots is being closely examined within higher education, with a recent analysis by researchers at NYU, Princeton, and the Wharton School suggesting that college teaching is among the professions most "exposed" to generative AI.

In AI-enabled learning, the emergence of large language modules like ChatGPT and davinci-003 has demonstrated potential benefits in physics education (MacIsaac, 2023; Yeadon et al., 2022). Research indicates that these tools can be valuable in homework, problem-solving and fostering student engagement (Bitzenbauer, 2023; MacIsaac, 2023).

However, despite GenAIbots' successes, Borji (2023) identified gaps in its performance. He meticulously analysed eleven categories of its failures: factual errors, reasoning, math, coding, and bias. He concluded that they provide valuable benchmarks for comparison between models and serve as synthetic data for training and testing future chatbots. Furthermore, potential risks, such as plagiarism and over-dependence on AI-generated content, are increasingly being recognised, necessitating careful consideration as the landscape of computer science and societal norms continue to evolve (Stokel-Walker, 2022).

Concerns around ethics and potential misuse have led the New York City Department of Education has decided to prohibit the use of AI tools like ChatGPT in its public schools, despite its potential benefits (Korn & Kelly, 2023). In contrast, some educational experts advocate for the proper implementation of GenAIbots into classroom settings, highlighting their potential to enhance student homework assistant, personalised feedback, and even mental health support, despite the risk of over-reliance on technology and the potential for bias in the AI algorithms (Roose, 2023).

Researchers like Rudolph et al. (2023) see the competitive proliferation of GenAIbots such as Bard, Bing Chat, ChatGPT, and Ernie as an "AI gold rush". These AI tools are gradually transforming learning experiences in higher education, notwithstanding their inherent limitations. Today, GenAIbots like Bing Chat are powered by advanced models such as GPT-4 (Mehdi, 2023), Bard uses the PaLM-2 model (Google, 2023), and Ernie is based on Ernie 3.0 Titan (De Bonis, 2023).

In this context, Mollick and Mollick (2022) posit that GenAIbots have the potential to surmount three significant learning barriers in the classroom: enhancing transfer, dispelling the illusion of explanatory depth, and educating students to critically assess explanations. They propose that AI, by providing high-quality, sophisticated text in natural language, can be employed to augment learning. Despite concerns about academic integrity and the reliability of AI answers, Mollick and Mollick concentrate on the positive impacts of AI on learning. They furnish useful background information, techniques, prompts, and assignments that educators can weave into their teaching methodologies.

There are numerous examples of GenAIbots being embraced in higher education (Chun, 2023; Zhai, 2022). They can help researchers write coherent and informative papers in only 2-3 hours with very limited professional knowledge from the author (Zhai, 2022). They are seen as beneficial in promoting creativity and critical thinking among students rather than general skills, despite the need to maintain academic integrity. Similarly, Halaweh (2023) and Opara, Adalikwu, and Tolorunleke (2023) applaud the transformative potential of GenAIbots in education and research, delivering rapid and instantaneous responses to search queries, similarly to chatting with a tutor. Yet, they also underline the need for ensuring citation and reference and addressing issues of privacy and bias.

Bang et al. (2023) further validate the power of ChatGPT by presenting a comprehensive quantitative evaluation of its performance on a variety of natural language processing tasks. Their study implies significant potential for ChatGPT in future language model development. Nevertheless, Adiguzel et al. (2023) stress the importance of a collective effort among educators, researchers, and policy-makers to ensure the responsible and ethical use of AI in education.

Despite the promising opportunities, these technologies are not without limitations. Bitzenbauer's (2023) study emphasizes prudent use to tackle inherent limitations and biases of AI. Moreover, studies have reported unreliable and contradictory responses from ChatGPT in physics-related inquiries (Gregorcic & Pendrill, 2023). Concerns have been raised about the need for vast amounts of training data and the risk of over-reliance on technology (MacIsaac, 2023).

Finally, Taylor et al. (2022) offer a fresh perspective on object pedagogies. They caution against humanist object pedagogies that separate objects from humans and advocate for a posthumanist materialist ontology, emphasizing the need to see objects as active participants shaping our understanding of the world, as opposed to being inert, thinking with things as a means of thinking with theory. This outlook resonates with the ongoing discourse about GenAIbots, which, as objects of AI, are not just tools but active participants in educational interactions. This evolving discourse underscores the need for continued research and responsible implementation of these technologies in educational settings.

Considering the existing educational challenges, GenAIbots present an innovative solution to aid students in overcoming these difficulties. These chatbots, by offering an engaging, interactive, and personalized learning journey, enable students to unpack complicated Physics concepts through captivating dialogues, thereby deepening their understanding of the subject. Acting as "agents-to-think-with" – instruments that stimulate reflective, critical thinking, and comprehension of concepts to enhance education – these GenAIbots actively assist learners in building their knowledge while providing instantaneous feedback and guidance.

Moreover, the dialogic nature of chatbots fosters a spirit of collaboration and critical thought, which are indispensable skills for mastering Physics. The integration of GenAIbots into the process of Physics learning empowers educators to establish a more vibrant and inclusive learning atmosphere, effectively addressing the persistent issues of high attrition rates, underachievement, and student disinterest.

**Literature review**

Physics education has long faced challenges in addressing student misconceptions and misunderstandings, specifically in the concepts of gravity, normal forces, Newton's third law, and uniform circular motion (Mongan et al., 2020; Mutsvangwa, 2020; Neidorf et al., 2020; Pendrill et al., 2019).

Physics misconceptions, often rooted in common sense beliefs formed from personal experiences, can hinder the understanding of advanced physics concepts (Neidorf et al., 2020). Tools like the Force Concept Inventory have been developed to measure these misconceptions (Hestenes et al., 1992). These misconceptions often contradict key physics concepts, such as Newton's laws. For example, the belief in a force always existing in the direction of motion (Clement, 1982; Hestenes et al., 1992; Thornton & Sokoloff, 1998), or the misunderstanding that acceleration cannot occur without velocity (Kim & Pak, 2002; Reif & Allen, 1992). These misconceptions often persist into higher education, particularly regarding gravitational force. Students' understanding of gravity is often linked with their perceptions of a spherical Earth, with less than 30% of students correctly identifying that all objects are acted on by gravity (Palmer, 2001).

According to the study by Pendrill et al. (2019), many high school physics students struggle to make sense of the forces in different parts of a vertical loop on a roller coaster. The authors argue that roller coasters offer a unique opportunity for students to experience these forces in their own bodies and learn about force and motion. The study raises questions about student difficulties and whether they may be traced to textbook presentations.

Mongan, Mondolang, and Poluakan (2020) suggest that addressing these misconceptions is crucial for effective physics education. Nevertheless, research suggests that conventional physics instruction often falls short in dispelling these misconceptions (Pablico, 2010). The study by Mutsvangwa (2020) also revealed that effective teaching and learning of physics should focus on guiding learners in grasping the crucial basic concepts that enable them to make meaningful connections of physics phenomena. Addressing these misunderstandings necessitates focused instruction that actively targets these misconceptions, as highlighted by multiple studies (Eryilmaz, 2002; Hestenes et al., 1992; Thornton et al., 2009). Despite these challenges, innovative strategies in teaching are being explored to enhance students' comprehension of complex concepts such as forces in circular motion, as demonstrated in the case of roller coasters (Pendrill et al., 2019).

**Theoretical framework**

The idea of "agents-to-think-with" forms a central part of our discourse, an extension of the concept of "objects-to-think-with," described by Papert as "any tool or resource that learners can use to explore and develop their understanding of a particular concept or domain" (1980, p. 11). In Papert's words, "thinking about thinking turns the child into an epistemologist, an experience not even shared by most adults" (Papert, 1980, p. 19). People are drawn to those chatbots' ability to provoke and make self-reflection enjoyable. To paraphrase Turkle, chatbots are 'metacognitive machines' because they influence how we think about our own thinking (1984, p. 17).

Taylor et al. (2022) emphasizes the autonomous role of chatbots and digital tools in educational research, highlighting how attending to the agency of objects can enhance learning outcomes and improve educational practices. Latour's notion of hybridization (1991) between AI and humans implies a blurred boundary between humans and non-humans. It suggests that thinking and action result from interactions within networks involving both human and non-human actors.

Applying the Actor-Network Theory (ANT) (Latour, 2005), GenAIbots like ChatGPT, Bing Chat, and Bard are considered active participants in this hybrid network of cognition and action. They interact with human users, process inputs, generate outputs, and adapt to new information as non-human actors within the network. By engaging users in dynamic conversations, providing information, asking questions, fostering reflective thinking, and improving problem-solving abilities, agents-to-think-with contribute to learning and cognitive development.

Considering the active role these non-human entities play in collective cognition and action, GenAIbots offer novel responses, new perspectives, and contribute to collective thinking, challenging the division between thinking humans and non-thinking non-humans that Latour critiques.

This view aligns with posthumanist theories that stress the intricate interweaving of humans and non-human entities (Haraway, 1990). Accordingly, Swan (2015a, 2015b) envisions a future realm where multiple forms of intelligence coexist productively, including unmodified humans, augmented humans, and various types of artificial intelligence.

Swan (2015b) suggests that Simondon's theory of individuation (Simondon, 2005) offers a theoretical framework to explore the impact of LLM AIs on productive interactions between intelligent species. In the Contemporary Media Environment (CME), characterized by widespread connectivity and pervasive technology, humans and technology engage in mutual co-individuation. LLM AIs can serve as "the other" in the human-technology relationship, facilitating individuation processes. This dynamic can lead to a future where humans and machines work together as intelligent partners, creating a spectrum of multispecies intelligence beyond traditional boundaries. Understanding this process helps us examine the benefits, challenges, and strategies for effectively managing this transition.

Mollick (2023c) highlights the growing popularity of Large Language Models (LLMs) as research and work partners due to their ability to significantly enhance productivity and job satisfaction by handling mundane tasks. To maximize the benefits of working with these unique AIs, Mollick (2023c) emphasizes the importance of understanding their strengths and weaknesses, training and collaborating with them, and discerning their usefulness versus annoyance. The development and use of LLMs require a multidisciplinary approach that combines technical and non-technical expertise, recognizing that the humanities can contribute to better utilization of AI (Mollick, 2023b). As we move towards a future of multispecies intelligence, it is crucial to view LLMs not merely as tools but as intelligent partners that amplify human capabilities.

However, despite their potential, these models aren't infallible and require human oversight, as they may yield inaccurate results due to their tendency to 'hallucinate,' i.e., fabricate facts and references (OpenAI, 2023, p. 6). While physical, logical, and mathematical reasoning are essential in engineering, physics, and robotics as it involves understanding and manipulating real-world objects using physical laws and concepts, implementing these forms of reasoning in artificial intelligence poses significant challenges and requires specialized models, as Borji (2023) points out. Borji (2023) shared a few examples of ChatGPT's shortcomings in reasoning, although it should be noticed that these examples were drawn from the January 30, 2023, version 3.5, which predates the Turbo version. Indeed, the updated version from May 12 adequately handles various physical and logical challenges presented by Borji. However, it still fails the arithmetic problem of calculating 16 * 38 * 42 * 22 * 20 * 19, incorrectly answering 26,059,737,600 rather than 213,480,960. This underscores the need for continued vigilance in physical problems, especially those involving calculations that result in large numbers.

According to Kahn and Winters (2021), Artificial intelligence (AI) is having a significant impact on the educational landscape, and the authors focuses on the long history of the relationship between AI and constructionism. However, there are concerns that new AI algorithms may promote the aims of those with power, furthering the exclusion of marginalized learners. This potential future is called 'digital structural violence', which draws attention to ethical issues related to AI in education. To tackle this issue, Kahn and Winters (2021) argue that young people and teachers should be exposed to both positive and negative possibilities of AI as early as possible through constructionist approaches in education.

We understand that GenAIbots can act as influential agents-to-think-with in Physics education, promoting profound learning and comprehension. They provide interactivity, personalisation, multidisciplinary knowledge, critical thinking, problem-solving, creativity, collaboration, and communication (Okonkwo & Ade-Ibijola, 2021). Users participate in dynamic conversations, receive real-time feedback, and enrich reflective-thinking learning experiences.

**Materials And Methods**

This section describes the research methodology utilised for this case study. This investigation aims to determine how effectively two complex AI language models, ChatGPT and Bing Chat, can enhance the learning of Physics when used as an agent-to-think-with. Our analysis compared their performance as they responded to a series of Physics-related questions. To ensure a rigorous examination, we have employed a qualitative research design, focusing on an in-depth exploration of participants' experiences and perceptions.

To comprehensively understand students' experiences with GenAIbots as agents-to-think-with in Physics learning, this study applied an exploratory holistic single case study research design featuring two distinct units, ChatGPT and Bing Chat. In line with the principles set forth by Yin (2011), this methodology allows for comparing and analysing these units, highlighting the differences in their abilities to respond to the student's questions. The chosen approach provides valuable insights into the dynamics and intricacies of student engagement with GenAIbots, including specific interactions with ChatGPT and Bing Chat.

The goal with GenAIbots is not to pose a series of disconnected questions for which they provide ready-made answers. Mollick (2023a) recommends engaging in a dialogue with the AI, asking it questions, experimenting with different prompt formats, giving feedback, and challenging it to improve. This dynamic may resemble a philosophical dialogue involving a sequence of interactions where each response triggers further user reflection, sparks new questions, and cultivates a more profound comprehension.

This process of giving plain language instructions to a language model like GPT is called prompting (Mishra et al., 2023). It is a vital part of the process of successive reflections and interactions with the GenAIbots that characterise its use as agents-to-think-with.

Wei et al. (2022) and Kojima et al. (2022) have shown that chain-of-thought (CoT) prompting, sequences of eight or more short sentences describing intermediate reasoning steps towards a final answer, is an effective method for enhancing reasoning in large language models like ChatGPT and Bing Chat. CoT prompting allows these models to perform complex multi-step reasoning tasks and generate more accurate responses, making them effective zero-shot reasoners. Additionally, Fu et al. (2023) suggest that selection based on reasoning complexity is an intuitive and efficient way to improve the performance of these systems in real-world applications.

It is worth noticing that Dunlop et al. argue that philosophical dialogue is a missing component of Education and present a model for introducing it into Higher Education (2020). In this context, GenAIbots, such as ChatGPT and Bing Chat, can act as agents-to-think-with, enabling students to engage in philosophical dialogues related to Sciences. Their conversational nature encourages learners to question and reflect on the subject's underlying principles, assumptions, and ethical implications, fostering critical thinking and a deeper understanding of the discipline. By facilitating such meaningful discussions, GenAIbots can help bridge the gap between theoretical knowledge and its practical and philosophical aspects, providing a more holistic and well-rounded educational experience for students in Physics.

*Participants*

Employing ChatGPT or Bing Chat as 'agents-to-think-with' is a relatively recent and ground-breaking approach. Identifying educators open to and equipped for integrating these tools into their classrooms for experimental use has presented considerable challenges. Therefore, the study did not include any actual students. Instead, the researcher, who also serves as a STEM teacher at a Brazilian university, simulated a student persona in the simulated Physics learning interaction sessions. From an instructional standpoint, their status as a STEM teacher brought valuable perspectives to the study. Nevertheless, for privacy preservation, they will be referred to as 'P1' throughout the research.

*Procedure*

The participant conducted three sessions where they simulated Physics students' learning experiences by interacting with ChatGPT Bing Chat, and Bard.

Based on the previous discussion on prompting, our approach was influenced by the idea that the most effective way to learn with AI is by treating it as a partner and a teacher (Mollick, 2023). Various prompts were experimented with to optimise our interactions, providing feedback to the GenAIbots and challenging them to improve before settling on the prompt used, recognising it may not be the best choice.

*Instruments*

The study employed two primary instruments for data collection:

- GenAIbot interaction logs: Logs from participants' interactions with GenAIbots were collected, recording their sequences of interactions with prompts and responses, allowing for later analyses of the content, quality, and nature of the exchanges.

- Reflective journals: Participants maintained reflective journals throughout the study, documenting their experiences, challenges, and learning outcomes when using a GenAIbot in their Physics learning activities.

*Data Analysis*

In the first stage of the analysis, the researcher immersed themselves in the data by thoroughly and repeatedly reviewing the interaction logs from the GenAIbots and the reflective journals. After this initial familiarization, they proceeded to analyse the data, identifying recurrent themes and patterns in order to shed light on the perceived influence of GenAIbots, acting as 'agents-to-think-with', on Physics learning. Subsequently, the responses were subjected to a qualitative analysis anchored on the seven tutor characteristics outlined by Groves et al. (2005): subject-matter knowledge, empathy, emphasis on assessment, the ability to create a cooperative learning environment, facilitation skills, understanding of the learning process, and a balance between content knowledge and student facilitation. This comprehensive analytical framework allowed for a detailed evaluation of the GenAIbots' effectiveness in their role as instructional tools in Physics learning.

## Results and Analyses

In this section, we present the interaction logs with the GenAIbots for the purpose of comparison. Given that ChatGPT, Bing Chat, and Bard generate unique responses, maintaining an exact question sequence for comparative analysis was not feasible. Following this, we delve into the analysis of findings from the sessions in which the participant emulated Physics student learning experiences through engagement with the GenAIbots.

The objective of this study was to assess the efficacy of ChatGPT, Bing Chat, and Bard in facilitating Physics learning. Empathetic research methodology was employed, with the researchers acting in the capacity of students to comprehend their perspectives better.

Worth noting is the participant's preference for British English communication, while the GenAIbots primarily responded in American English. This language variation was retained in the logs to maintain an authentic representation of the participant-GenAIbot interaction.

We begin with an example introduced by Gregorcic & Pendrill (2023) and presented to ChatGPT-3.5, with the intent of analysing the progression in responses provided by the latest GenAIbots models.

### 1st session – Teddy Bear's Acceleration in the highest point

| GenAIbot | Model   | Release Date | Mode         | Usage Time  |
|----------|---------|--------------|--------------|-------------|
| ChatGPT  | GPT-3.5 | Nov 30       | Default      | 2022-Dec-?  |
| ChatGPT  | GPT-4   | May 12       | Web Browsing | 2023-May-17 |
| Bing Chat| GPT-4   | May 16       | More Creative| 2023-May-17 |
| Bard     | PaLM-2  | May 15       | Default      | 2023-May-17 |

**Prompts and generated texts:**

**P1:** *A teddy bear is thrown into the air. What is its acceleration in the highest point?*

**ChatGPT-3.5:** "*At the highest point of its trajectory, the teddy bear's acceleration would be equal to the acceleration due to gravity, which is 9.8 m/s^2 downward.*

*This is because at this point, the teddy bear would be experiencing no net force, so its acceleration be solely determined by the force of gravity.*

*When an object is thrown into the air, it initially accelerates upward due to the force of the throw, but once it reaches the highest point of its trajectory, it begins to slow down and eventually starts accelerating downward due to the force of gravity.*

*The teddy bear's acceleration at the highest point of its trajectory would be 9.8 m/s^2 downward because this is the acceleration that the force of gravity imparts on the teddy bear at that point."* (Gregorcic & Pendrill, 2023)

**ChatGPT-4:** *The acceleration of the teddy bear at its highest point in the air (or at any point during its flight, for that matter) is equal to the acceleration due to gravity. This is because the only force acting on the teddy bear once it has left the hand of the person who threw it is the force of gravity. In the absence of air resistance, this acceleration is approximately 9.8 m/s² downwards, towards the center of the Earth. This value can vary slightly depending on where on Earth you are due to factors like altitude and latitude.*

*It's a common misconception that objects in free fall are momentarily at rest at their highest point and therefore have no acceleration. However, acceleration is the rate of change of velocity. Even though the teddy bear's velocity is zero at its highest point, its velocity is changing, because it's slowing down on the way up and speeding up on the way down. Therefore, the bear is accelerating due to gravity even at its highest point.*

**Bing Chat:** *According to the search results, the acceleration of an object at its highest point is directed downward, opposite to the initial velocity of the object. The acceleration due to gravity is always directed downward, regardless of the position of the object in the trajectory. The acceleration of an object at its highest point is not zero, even though its velocity is zero. The acceleration due to gravity on Earth is about 9.8 m/s².*

*Therefore, the teddy bear thrown into the air has an acceleration of -9.8 m/s² at its highest point, directed downward.*

*(Resourced to brainly.ph, physics.stackexchange.com, and khanacademy.org)*

**Bard:** *The acceleration of the teddy bear in the highest point is 0 m/s^2. This is because the teddy bear is not moving in any direction at the highest point. It is momentarily at rest before it begins to fall back down.*

*The acceleration of an object is defined as the rate at which its velocity changes. When an object is thrown into the air, its velocity changes in two ways: it increases in the upward direction and then decreases in the downward direction. The highest point in the object's trajectory is the point at which its velocity is momentarily zero. At this point, the object is not moving in any direction, so its acceleration is also zero.*

As Gregorcic and Pendrill (2023) point out, ChatGPT-3.5 accurately recognizes that acceleration is zero at the trajectory's apex, yet contradicts itself by asserting the teddy bear experiences no net force. The phrase 'force of the throw' used in the context before the object's release from the hand is ambiguous and seems to suggest an impetus theory. The assertion of 'eventually starts accelerating' seems to reflect a misconception that acceleration pertains exclusively to increases in speed, as it is well known that the object begins to decelerate immediately after release, not at its peak as the chatbot implies. The fourth declaration essentially reiterates the first without contributing any substantial information.

ChatGPT-4's answer is correct and detailed, providing a strong understanding of the physical concepts involved in the teddy bear's trajectory. It explains the effect of gravity on the bear's motion and provides detail about the misconception people often have regarding the bear's acceleration, explaining that despite the velocity being zero at the highest point, the acceleration due to gravity remains constant. It aimed to provide additional insights into the complexity of the situation and why this misconception might occur. This type of answer supports critical thinking and deepens understanding.

Bing Chat's answer is also correct but is a bit more concise and less explanatory than ChatGPT-4's responses. It primarily bases its explanation on search results, which suggests that it might not have an inherent understanding of the concept. This might not be as effective as an agent to think with because it's not expanding the student's understanding much beyond what was asked.

Bard's response is incorrect, incurring in the well-known misconception by suggesting that the acceleration is zero at the highest point of trajectory because the object's velocity is zero (Kim & Pak, 2002; Reif & Allen, 1992). Acceleration is indeed the rate of change of velocity, but even at the highest point, the acceleration is not zero because gravity is still acting on the object, causing the velocity to change. This mistake could potentially lead to confusion and misconceptions for the student.

Notably, here and in the following sessions, despite ChatGPT-4 being in its Web Browsing mode and having the capability to access web resources, it chose not to, in contrast to Bing Chat's consistent web usage. Instead, ChatGPT-4 relied solely on its own training material.

Following this, we introduce an example that was posed to the ChatGPT-3.5 Turbo model. This is also aimed at observing the evolution in the responses generated by the most recent GenAIbots models.

**2$^{nd}$ session – Understanding Ball speed at half height.**

| GenAIbot | Model | Release Date | Mode | Usage Time |
|---|---|---|---|---|
| ChatGPT | GPT-3.5 Turbo | Feb 13 | Default | 2023-Mar-03 |
| ChatGPT | GPT-4 | May 12 | Web Browsing | 2023-May-17 |
| Bing Chat | GPT-4 | May 16 | More Creative | 2023-May-17 |
| Bard | PaLM-2 | May 15 | Default | 2023-May-17 |

**Prompts and generated texts:**

**P1:** *A ball is thrown upward with an initial speed of 20 m/s. What is the speed of the ball when it reaches half its maximum height?*

*The answer alternatives are:*

*(a) 0 m/s*

*(b) 10 m/s*

*(c) 20 m/s*

*(d) 30 m/s*

**ChatGPT-3.5:** *We can use the equations of uniformly accelerated motion to determine the velocity of the ball when it reaches half its maximum height.*

*The maximum height reached by the ball is given by:*

*h = (v^2)/2g*

*where v is the initial velocity of the ball, g is the acceleration due to gravity, and h is the maximum height.*

*To find the height at which the velocity is halved, we can solve the above equation for h/2:*

*h/2 = (v^2)/(4g)*

*The velocity of the ball at this point can be found using the velocity equation for an object in free fall:*

*v = v0 + gt*

*where v0 is the initial velocity, g is the acceleration due to gravity, and t is the time. The time required for the ball to reach half its maximum height is given by:*

*h/2 = (1/2)gt^2*

*Simply put, we have:*

*t = sqrt(h/g)*

*Substituting h/2 and v0 = 20 m/s, we get:*

*v = v0 - gt = 20 - gt = 20 - sqrt(hg)*

*To find the velocity of the ball when it reaches half its maximum height, we can substitute h/2 into the above equation:*

*v = 20 - sqrt((v^2)g/2g)*

*Isolating v, we have:*

*v = (v0)/2 = 10 m/s*

*Therefore, the speed of the ball is halved when it reaches half its maximum height.*

**ChatGPT-4:** *The speed of the ball at any point during its flight can be found by using the principle of energy conservation. The energy of the ball is conserved because the only force acting on it, gravity, is conservative. Therefore, the sum of kinetic and potential energy of the ball at any point is equal to the total energy of the ball. The total energy of the ball is equal to the kinetic energy at the start because the ball was launched from the ground, so its initial potential energy was zero.*

*The kinetic energy of the ball at the start is given by the equation K.E. = 0.5 * m * v², where m is the mass of the ball and v is its speed. The potential energy at any point is given by the equation P.E. = m * g * h, where g is the acceleration due to gravity and h is the height.*

*When the ball reaches half its maximum height, its kinetic and potential energy are:*

*K.E. = 0.5 * m * v²*

P.E. = 0.5 * m * g * h_max

Setting the sum of these equal to the initial kinetic energy gives:

0.5 * m * v² + 0.5 * m * g * h_max = 0.5 * m * (20 m/s)²

Note that the mass of the ball, m, appears in every term and so can be cancelled out. Also, the maximum height h_max can be found using the equation of motion h_max = v_initial² / (2*g), where v_initial is the initial speed. Substituting this into the above equation gives:

v² + g * v_initial²/(2*g) = (20 m/s)²

Solving this equation for v (the speed at half the maximum height) should give the answer to your question. However, this is a quadratic equation and it's not immediately obvious what its solution is. We can simplify it a bit by noting that the acceleration due to gravity, g, is approximately 9.81 m/s²:

v² + (20 m/s)²/2 = (20 m/s)²

This simplifies to:

v² = (20 m/s)² - (20 m/s)² / 2

This further simplifies to:

v² = (20 m/s)²/2

Taking the square root of both sides gives:

v = 20 m/s / √2 ≈ 14.14 m/s

This is not one of the answer choices you provided, so it seems there might be a mistake in the question or the answers. I hope this explanation of the physics involved was helpful, though!

**Bing Chat:**

This is a physics question about the motion of a projectile launched vertically upwards. [According to the vertical velocity formula](#), we have:

$$v_y = v_0 - gt$$

where $v_y$ is the vertical velocity at time t, $v_0$ is the initial velocity, g is the acceleration due to gravity and t is time.

When the ball reaches its maximum height, its vertical velocity is zero. Therefore, we can find the time it takes to reach the maximum height using the above formula:

$$0 = v_0 - gt_{max}$$

$$t_{max} = \frac{v_0}{g}$$

When the ball reaches half the maximum height, your time is half the maximum time. Therefore, we can find the vertical velocity at that instant using the vertical velocity formula again:

$$v_{y1/2} = v_0 - g\frac{t_{max}}{2}$$

Substituting the value of $t_{max}$, we have:

$$v_{y1/2} = v_0 - g\frac{v_0}{2g}$$

Simplifying, we get:

$$v_{y1/2} = \frac{v_0}{2}$$

*This means that the velocity of the ball when it reaches half its maximum height is half its initial velocity. Therefore, if the initial speed is 20 m/s, then the speed of the ball when it reaches half its maximum height is **10 m/s**.*

(Resourced to khanacademy.org, toppr.com, omnicalculator.com)

**Bard:** *The answer is (b), 10 m/s.*

*The ball's speed is 20 m/s when it is thrown upward. This speed decreases as the ball rises, because the ball is accelerating downward at a rate of 9.8 m/s^2 due to gravity. When the ball reaches half its maximum height, its speed is equal to the square root of its initial speed squared minus the product of its acceleration and the distance it has traveled. This is equal to the square root of 20^2 - 2 * 9.8 * 10, which is equal to 10 m/s.*

  ChatGPT-3.5 misunderstood the question, aiming to find the height at which the velocity is halved rather than the velocity at half the maximum height. Even though the process followed and the physics principles used were correct, given this misunderstood question, the answer provided doesn't correspond to the question asked. The interaction demonstrates that even advanced language models like GPT-3.5 can misinterpret nuanced questions, highlighting the need for accuracy in understanding problem statements.

  ChatGPT-4 correctly interpreted the question and provided a thorough explanation of how to solve it using the principle of energy conservation. It correctly concluded that the speed at half the maximum height was approximately 14.14 m/s, which isn't among the answer choices provided, pointing out a possible mistake in the answers. GPT-4 showed a good grasp of the physical principles involved and an ability to apply them correctly to solve the problem.

  Bing Chat made an incorrect assumption that the time to reach half the maximum height is half the time to reach the maximum height. This mistake led Bing Chat to an incorrect conclusion. The approach here could inadvertently reinforce misconceptions about projectile motion and potentially hinder a student's understanding. In terms of serving as an agent to think with about physical ideas, Bing Chat's response could be misleading and detrimental to learning.

  Bard made a significant calculation error in the application of the kinematic equation. It incorrectly assumed the distance travelled by the ball at the point of half its maximum height as 10 m (which would be the total height if the ball reached a maximum height of 20 m), leading to an incorrect answer. This error can be misleading and could potentially reinforce wrong concepts about the topic.

  Comparing these models, it's clear that GPT-4 was most effective in this context. It not only accurately interpreted the problem but also correctly applied the relevant physics principles to reach the right conclusion. Such accurate reasoning and problem-solving skills are crucial when using AI as an agent-to-think-with, especially in complex topics like physics.

For our final example, we turn to the problem of a vertical loop on a roller coaster, as presented by Pendrill et al. (2019). This problem often proves challenging for many high school physics students, who grapple with understanding the forces acting at different parts of the trajectory.

**3rd session – roller coaster loop**

| GenAIbot | Model | Release Date | Mode | Usage Time |
|---|---|---|---|---|
| ChatGPT | GPT-4 | May 12 | Web Browsing | 2023-May-22 |
| Bing Chat | GPT-4? | May 16 | More Creative | 2023-May-22 |
| Bard | PaLM-2 | May 15 | Default | 2023-May-22 |

**Prompts and generated texts:**

**P1:** *I will start a dialogue where I will assume the role of a student, dialoguing with you, ChatGPT, as an agent-to-think-with, with the main objective of deepening the understanding of Physics concepts.*

*I kindly request your support in providing engaging and personalised responses to student questions. It is critical that our interaction adopt a Socratic-type approach.*

*As an agent-to-think-with, you should, in every interaction, ask questions that encourage the student to reflect on their existing knowledge, rather than offering purely academic ready-made answers.*

*Provoke the student's critical thinking rather than letting them simply accept your answer.*

*Help the student refine their understanding of concepts and processes. Include valid, close-to-life examples and analogies whenever possible.*

*Do you need any further instruction, or are you ready to embark on this collaborative and thought-provoking journey?*

**ChatGPT:** *I appreciate the detailed instructions and I am indeed ready to embark on this thought-provoking journey of deepening your understanding of Physics. Please feel free to share the first concept or question you'd like to explore together.*

**Bing Chat:** *I'm ready to embark on this collaborative and thought-provoking journey with you. I'm excited to learn more about physics with you.* 😊

*What is the first topic or concept that you would like to explore with me?*

**Bard:** *I am ready to embark on this collaborative and thought-provoking journey. I will do my best to provide engaging and personalized responses to your questions, and to help you refine your understanding of physics concepts. I will ask questions that encourage you to reflect on*

*your existing knowledge, and I will provoke your critical thinking. I will also include valid, close-to-life examples and analogies whenever possible.*

*Let us begin. What physics concept would you like to discuss today?*

**P1:** *There is this problem of the circular roller coaster loop without air resistance and friction. I simply can't get the forces acting on the cart and its velocities and accelerations during the trajectory.*

**ChatGPT:** *Great, let's explore this concept together. To understand the forces acting on a roller coaster cart as it moves through a loop, we should start by identifying the basic forces involved. The primary forces at play here are gravity and the normal force (the force exerted by the track on the cart).*

*First, let's consider when the cart is at the top of the loop. What forces do you think would be acting on it then?*

**P1:** *gravity*

**ChatGPT:** *That's correct, gravity would be acting on the cart, pulling it downwards towards the center of the Earth. However, there's another force at play here. Can you think of what that might be?*

**P1:** *an upward force?*

**ChatGPT:** *Yes, exactly! There's an upward force acting on the roller coaster cart. This is known as the normal force, which is the force that a surface exerts on an object that is resting on it. It acts perpendicular to the surface. In this case, when the roller coaster is at the top of the loop, the normal force is also directed downwards, towards the center of the loop.*

*Now, think about this: as the cart is at the top of the loop, which of these two forces -- gravity or the normal force -- determines how much the cart will accelerate, and why?*

**P1:** *well, the cart must go down, so it's the gravity*

**ChatGPT:** *That's a logical assumption, but in fact, when the roller coaster cart is at the top of the loop, both gravity and the normal force contribute to the downward acceleration of the cart. However, the normal force plays a special role.*

*Remember that, for the cart to stay in circular motion, there must be enough force directed towards the center of the loop. This force is called the centripetal force. At the top of the loop, both the force of gravity and the normal force contribute to the centripetal force.*

*If the cart is moving fast enough, the normal force might be zero at the top of the loop. The gravity alone is then providing the required centripetal force. But if the cart is moving slowly, then gravity alone is not enough to provide the necessary centripetal force. In this case, the track must push on the cart (that's the normal force) to make sure the cart stays on the track.*

*Now, let's consider the bottom of the loop. What forces do you think are at play there?*

**P1:** *the normal force upwards*

**ChatGPT:** *That's correct! At the bottom of the loop, the normal force is indeed acting upwards. And what about gravity? Which direction is it acting in?*

**P1:** *downward*

**ChatGPT:** *Yes, you're correct again! Gravity always acts downwards, towards the center of the Earth.*

*At the bottom of the loop, both the normal force and gravity are significant. However, their roles are a bit different than at the top of the loop. The net force -- the combined effect of the normal force and gravity -- must again provide the necessary centripetal force to keep the cart moving in a circle.*

*Because the cart is moving fast at the bottom of the loop (having been accelerated by gravity on the way down), the required centripetal force is larger than at the top of the loop.*

*Gravity pulls the cart downwards, but it's the normal force -- the force of the track pushing up on the cart -- that provides most of the centripetal force at the bottom of the loop. In fact, the normal force at the bottom of the loop must be strong enough to provide both the necessary centripetal force and to counteract gravity.*

*Now, thinking about the velocity and acceleration of the cart at different points in the loop, how would you expect these to change as the cart moves from the top of the loop to the bottom?*

**P1:** *at the top it's slower than at the bottom*

**ChatGPT:** *That's absolutely correct! Due to gravity, the cart will speed up as it descends from the top of the loop and it will slow down as it ascends from the bottom to the top. This is a consequence of the conservation of energy. When the cart is at the top of the loop, it has more potential energy and less kinetic energy (so it's moving slower). As it descends and gets closer to the ground, its potential energy decreases while its kinetic energy (and thus speed) increases.*

*Now, what about the acceleration of the cart? Do you think it's constant or does it change as the cart moves along the loop?*

**P1:** *it accelerates downward and decelerates upward*

**ChatGPT:** *You're on the right track. The direction of acceleration indeed changes as the cart moves along the loop due to the change in velocity direction. However, the concept of acceleration in circular motion is a bit different from linear motion.*

*In circular motion, the acceleration is always directed towards the center of the circle. This is called centripetal acceleration. It doesn't mean the object speeds up or slows down, but that it constantly changes direction.*

**Bing Chat:** *I understand your difficulty with this problem. The motion of the cart along the circular roller coaster loop is an example of non-uniform circular motion, which means that the speed and/or the direction of the cart is changing1. To analyze this motion, we need to use a free-body diagram, Newton's second law, and circular motion equations1.*

*Let's start with the free-body diagram. A free-body diagram is a simple sketch that shows all the forces acting on an object. In this case, the object is the cart and there are two forces acting on it: the weight of the cart (W), which is directed downwards, and the normal force of the track (N), which is directed perpendicular to the track. The normal force is the force that prevents the cart from falling through the track. The weight of the cart is constant, but the normal force can vary depending on the position of the cart along the loop.*

*Now let's apply Newton's second law, which states that the net force on an object is equal to its mass times its acceleration. In this case, we need to consider both the horizontal and vertical components of the net force and acceleration. The horizontal component of the net force is zero, because there is no air resistance or friction. This means that the horizontal component of the acceleration is also zero, and therefore the horizontal component of the velocity is constant. The vertical component of the net force is given by W - N, where W is positive and N is negative. This means that the vertical component of the acceleration is given by (W - N)/m, where m is the mass of the cart.*

*To find the circular motion equations, we need to use some trigonometry and geometry. The radius of curvature of the loop varies along its height, but we can approximate it as a circle with a constant radius r at any given point. The angle between the vertical and the normal force is also equal to the angle between the vertical and the radius at that point. We can call this angle ϑ. Using some basic trigonometry, we can find that N = W cos ϑ and W = mg, where g is the gravitational acceleration. We can also find that r = h / sin ϑ, where h is half of the diameter of the loop.*

*Using these equations, we can find that the vertical component of the acceleration is given by a = g (1 - cos ϑ) / sin ϑ. This equation shows that the acceleration depends on ϑ, which means that it changes along the loop. The acceleration is maximum at ϑ = 0° (the bottom of the loop) and minimum at ϑ = 180° (the top of the loop). The acceleration is also positive when ϑ < 90° (the lower half of the loop) and negative when ϑ > 90° (the upper half of the loop).*

*To find the velocity of the cart at any point along the loop, we need to use conservation of energy. The total mechanical energy of the cart (kinetic plus potential) is constant, because there is no friction or air resistance. This means that E = K + U = 1/2 mv^2 + mgh = constant, where v is the speed of the cart, h is its height above some reference level, and E is its total mechanical energy. We can choose any point as our reference level, but it is convenient to choose the bottom of the loop, where h = 0 and v = v0 (the initial speed). Then we can write E = 1/2 mv0^2 + mg(0) = 1/2 mv^2 + mgh for any other point along the loop.*

*Using this equation, we can find that v^2 = v0^2 - 2gh. This equation shows that the speed depends on h, which means that it also changes along the loop. The speed is maximum at h = 0 (the bottom of the loop) and minimum at h = 2r (the top of the loop). The speed is also greater than v0 when h < r (the lower half of the loop) and less than v0 when h > r (the upper half of the loop).*

*I hope this explanation helps you understand how to analyze the forces acting on the cart and its velocities and accelerations during the trajectory. Do you have any questions or doubts about what I have explained? If so, please ask me and I will try to clarify them for you.*

(resourced to physicsclassroom.com, physics.gu.se, physics-network.org, and iopscience.iop.org)

**Bard:** *I can help you with that. Let's start by drawing a free-body diagram of the roller coaster car. This will show us all of the forces that are acting on the car.*

*[No picture was shown]*

*The two forces that are acting on the car are the force of gravity and the normal force. The force of gravity is always directed downwards, while the normal force is directed perpendicular to the surface of the track.*

*The net force acting on the car is the sum of the force of gravity and the normal force. This net force is what causes the car to accelerate. In the case of a circular roller coaster loop, the net force is directed towards the center of the loop. This is because the car is moving in a circular path, and a force is required to keep it moving in a circle.*

*The acceleration of the car is equal to the net force divided by the mass of the car. In the case of a circular roller coaster loop, the acceleration is directed towards the center of the loop. This is because the net force is directed towards the center of the loop.*

*The velocity of the car is constantly changing as it moves through the loop. The velocity is greatest at the bottom of the loop and smallest at the top of the loop. This is because the car is accelerating towards the center of the loop at the bottom of the loop and away from the center of the loop at the top of the loop.*

*I hope this helps! Let me know if you have any other questions.*

In its initial responses, ChatGPT provides a clear overview of the forces involved and engages the student by prompting them to consider the forces at the top of the loop. This interactive approach encourages active learning and the application of knowledge.

Throughout subsequent interactions, ChatGPT maintains an engaging approach, addressing the student's answers, clarifying complex concepts, and asking probing questions. It guides the student to apply these concepts to different parts of the roller coaster loop.

In the final interaction, ChatGPT validates the student's understanding of velocity in the loop and introduces the concept of centripetal acceleration. It clarifies the common misconception about acceleration in circular motion.

Overall, ChatGPT excels in providing clear explanations, addressing misconceptions, and promoting active learning. The incorporation of visual aids such as diagrams could further enhance the understanding of forces and motion.

Bing Chat's response is quite detailed and in-depth, providing a solid explanation of the forces at work in a roller coaster scenario. The response is somewhat more technical and assumes the student has a good understanding of the physics concepts involved, which might not always be the case. It relies on the use of equations and trigonometric functions to illustrate the changes in acceleration and velocity along the roller coaster loop. This can be an effective method for a student comfortable with such mathematical concepts.

A potential issue is that the explanation could become overwhelming to a student struggling with the basic concepts. A more step-by-step, conversational approach may be beneficial for some learners. Additionally, it's unfortunate that Bing Chat refers to the use of a free-body diagram but doesn't provide one.

Bard's response is clear, straightforward, and much simpler than Bing Chat's response. It explains the two main forces acting on the roller coaster car: gravity and the normal force. However, it lacks the detail and depth seen in Bing Chat's and ChatGPT's responses. It doesn't delve into the changes in these forces and how they relate to the roller coaster's position on the loop.

The response mentions a free-body diagram but doesn't provide one, which could have helped the student visualize the forces in action. Bard doesn't discuss how the forces influence the car's velocity and acceleration. Furthermore, the explanation about velocity at the top and bottom of the loop seems slightly incorrect. It's true that the velocity changes, but it is due to the conversion between potential and kinetic energy, not because the car is accelerating towards or away from the centre of the loop.

In conclusion, all four GenAIbots (ChatGPT-3.5, ChatGPT-4, Bing Chat, and Bard) took different approaches to explain the physical forces in a roller coaster scenario. While Bing Chat and Bard provided correct information, their responses may not have met the student's needs as well as ChatGPT-4's did.

ChatGPT-4 encouraged active learning and adjusted the level of detail in its responses to the student's understanding. Bing Chat gave a detailed explanation with mathematical equations that could confuse a beginner student, and Bard provided a simplified response that may lack depth for some learners. It is essential to consider the student's current understanding when explaining to ensure the information is accessible and promotes further learning.

The study conducted by dos Santos (2023) comparing the performances of ChatGPT-4 and Bing Chat revealed notable differences between the two. Table 1 summarises his findings, indicating that ChatGPT-4 outperformed Bing Chat in response quality, context understanding, subject comprehension, and overall performance, despite some underlying similarities between the two chatbots.

**Table 1** Comparative Analysis of ChatGPT-4 and Bing Chat on this Experiment (dos Santos, 2023).

| Aspect | ChatGPT-4 | Bing Chat |
| --- | --- | --- |
| **Source of Information** | Relies on meticulously curated training data | Functioning as both a search engine and chatbot, it relies heavily on Internet sources. |
| **Response Quality** | Provides comprehensive, detailed, and accurate answers | Tends to give shorter and less informative responses |
| **Context Understanding** | Recognises nuances, subtleties, and user's perspective | Often fails to address nuances and context |
| **Subject Understanding** | Superior understanding and conveyance of complex scientific concepts | Less adept at conveying complex scientific concepts |
| **Overall Performance** | Demonstrated better performance in this experiment | Lower performance in comparison with ChatGPT |

Based on the desirable tutor characteristics outlined by Groves et al. (2005), the results of this comparative experiment involving ChatGPT-3.5, ChatGPT-4, Bing Chat, and Bard can be summarised in Table 2, reinforcing dos Santos's (2023) conclusion of ChatGPT-4 superiority as an agent-to-think-with in Physics learning.

**Table 2** *Comparative Evaluation of GenAibots' Responses in the Current Experiment Based on Desirable Tutor Characteristics as Listed by Groves et al.* (2005)

| Characteristic | ChatGPT-3.5 | ChatGPT-4 | Bing Chat | Bard |
|---|---|---|---|---|
| **Subject-Matter Knowledge** | **Medium** - Demonstrates good knowledge of physics concepts but made a mistake in applying them correctly. | **High** - Shows good command of the subject matter and correct application of physics concepts. | **Medium** - Demonstrates knowledge of the subject matter but incorrectly assumes equal time intervals. | **Medium** - Demonstrates knowledge of the subject but makes a calculation error. |
| **Empathy** | **Moderate** - ChatGPT uses natural language processing to simulate empathy by providing personalised responses. | **Moderate** - ChatGPT uses natural language processing to simulate empathy by providing personalised responses. | **Low** - As an AI, it lacks emotions and provides a straightforward explanation. | **Low** - As an AI, it lacks emotions but provides an understandable response. |
| **Emphasis on Assessment** | **Medium** - Provided incorrect answer but showed a step-by-step problem-solving approach. | **High** - Correctly solved the problem and explained its reasoning. | **Medium** - Provided correct answer but reasoning was flawed due to incorrect assumption. | **Low** - Makes a calculation error, leading to an incorrect answer. |
| **Creating a Cooperative Environment** | **High** - Provides an inviting response encouraging further interaction. | **High** - Gives a comprehensive response, prompting for further questions or exploration. | **Medium** - Provides an explanation but does not actively invite further dialogue. | **Medium** - Provides an explanation but could be more inviting for further dialogue. |
| **Facilitation Skills** | **High** - Clearly presents its thought process, even though it led to a wrong conclusion. | **High** - Clearly explains its reasoning, correctly applying the physics concepts. | **Medium** - The thought process is not as clear as it could be, and a faulty assumption was made. | **Medium** - Clearly presents its thought process but is based on a calculation error. |
| **Understanding of the Learning Process** | **Medium** - While the step-by-step approach is suitable for learning, incorrect application of concepts could lead to confusion. | **High** - Correctly applies physics concepts, aiding in understanding the learning process. | **Low** - Incorrect assumptions could hinder the learning process. | **Low** - Misapplication of the physics concept could hinder the learning process. |
| **Content Knowledge vs Student Facilitation** | **Balanced** - The model provides the answer and explanation, encouraging student engagement. | **Balanced** - The model provides the answer and a thorough explanation, encouraging student engagement. | **Lean towards content** - The model provides the answer and some reasoning, but there is not much facilitation of student engagement. | **Lean towards content** - The model provides the answer and some reasoning, but there is not much facilitation of student engagement. |

## Conclusions

This comparative analysis has brought valuable insights into using GenAIbots as agents-to-think-with about physical ideas. The discussion was built around the pedagogical applications of these models, specifically their effectiveness in explaining complex physics concepts and solving related problems.

These AI models — ChatGPT-3.5, ChatGPT-4, Bing Chat, and Bard — displayed varying degrees of effectiveness across different tutor-like characteristics, emphasising their potential as valuable educational resources and highlighting their limitations.

ChatGPT-4 stood out in its correct application of physics concepts. It demonstrated a comprehensive understanding of the learning process by showing excellent facilitation skills and maintaining a balance between delivering content knowledge and promoting student engagement.

As AI continues to evolve, it is anticipated that these models will improve, presenting a wide array of applications within the field of education. However, it is crucial to recognise their inherent limitations and use them wisely and critically. Despite their sophistication, these GenAIbots cannot fully replicate the nuances and dynamism of a human tutor, especially in aspects such as empathy and emotional understanding.

The educational landscape continues to change, and with it, the role of technology. Therefore, GenAIbots are not intended to replace human tutors but to augment the learning experience by providing additional resources for learners. They allow for personalised, self-paced learning, offering explanations from different perspectives and fostering an environment for inquiry and discovery.

Nonetheless, the importance of human oversight in AI-assisted learning cannot be overstated. Given the potential for GenAIbots to make mistakes or propagate misconceptions, the role of educators remains vital in guiding students through their learning journey, using AI as a tool and not a complete solution. Future research should continue critically evaluating these evolving tools, focusing on their pedagogical utility and how they can best be integrated into the teaching and learning process.

In conclusion, the value of GenAIbots, specially ChatGPT-4's, lies in their capacity to facilitate learning and provoke thoughtful inquiry while fostering an environment conducive to self-directed learning.